# Study of interdomain boundary in diamagnetic domain structure in beryllium

Philip Lykov *RRC "Kurchatov Institute", 123182 Moscow, Russia*

November 21, 2002

**Abstract**

At low temperatures, in strong magnetic fields, the formation of a non-uniform magnetisation is possible in a single-crystal metal sample whose demagnetising factor along the field is close to unity. Namely, so-called Condon diamagnetic domain structure arises and disappears periodically with magnetic field. In this paper, the diamagnetic domain structure in beryllium single crystalis analysed. Directly, existence of diamagnetic domains in that sample was observed earlier by the muon spin precession (μSR) resonance peak splitting. A method is described that allows to calculate quantitative characteristics of the interdomain boundary using the muon histograms. The technique is based on the Marquardt minimisation procedure that has been modified in order to reduce the influence of noise on iterations convergence. Boundary volume fraction was calculated.

# 1 Introduction

The possibility of diamagnetic domain structure formation in a single-crystal metal sample was predicted in [1]. The domain structure does form, if the sample has demagnetising factor close to unity along the direction of external magnetic field, as it is shown in [1]. In such case, if the sample were uniformly magnetised the boundary condition for normal component of external magnetic field, $H_n$, would not be fulfilled due to the oscillatory field dependence of magnetisation. Experimentally, diamagnetic domains were for the first time observed by NMR in Ag [2]. In that paper it was shown that, a splitting occurred in NMR peak for external fields lying within some certain intervals. The positions of NMR doublet components indicated magnetic induction within the domains of either sort. After [2], for rather long a period, diamagnetic domain structure was out of scope of experimental study. It was only in 1995 that μSR technique was applied to study diamagnetic domain structure in beryllium single crystal [3]. Unlike NMR method that only allows to explore a thin skin layer, μSR method gives information of the magnetic field distribution within the bulk sample. The results of μSR experiments reported in [3] indicate unambiguously existence of diamagnetic domain phase. At the same time, the interdomain boundary is also of interest. The least information one can get from μSR histograms is the boundary volume fraction. Further details are heavily dependent on histogram time bin width and overall total.

# 2 Description of the method

The μSR method is based on correlation between the directions of muon spin and momentum vector of decay positron. A muon histogram is a set of values $N_k$, i.e. numbers of decay positrons, detected during the *k*-th time interval (so-called bin) within the range of directions limited by the detector space angle. The time interval for *k*-th bin is from $k \times \Delta t_0$ to $(k+1) \times \Delta t_0$, where $\Delta t_0$ is bin-width and *k* ranges from 0 to *N-1*. If we consider the experiment in that muons stop in a non-uniformly magnetised sample then the following approximation can be derived for $N_k$

$$\widetilde{N}_k = n_0\, e^{-\nu k}\, (1 + \alpha \int_{\beta_{min}}^{\beta_{max}} \frac{d\beta}{2\beta} (\cos(\beta\, n + \varphi_0) - \cos(\beta\,(n+1) + \varphi_0))\, f(\beta)) + b \qquad (1)$$

Here, $\beta \equiv \gamma\, \Delta t_0 B$ is dimensionless magnetic field, $\nu \equiv \Delta t_0/\tau_0$ - reduced bin-width, $\gamma = 85146.8069\, \text{s}^{-1}G^{-1}$ - gyromagnetic ratio of $\mu^+$, $\tau_0 = 2.197095 \cdot 10^{-6}\, s$ - muon lifetime, $\Delta t_0$ - bin-width ($0.625 \cdot 10^{-9}\, s$ in our case), *b* - background, $n_0 \equiv N_0\, \Delta t_0/\tau_0$ is proportional to the intensity of muon beam, $\alpha$ is the asymmetry coefficient, $\varphi_0$ is the initial phase determined by choice of moment $t = 0$. Equation (1) was derived with the assumption that $\Delta t_0 \ll \tau_0$. Values $\beta_{min}$ and $\beta_{max}$ determine possible range of magnetic fields in the sample and, in most general case, are equal to 0 and $+\infty$ respectively. The field distribution $f(\beta)$ is normalised so that

$$\int_{\beta_{min}}^{\beta_{max}} f(\beta)\, d\beta = 1 \tag{2}$$

In case of diamagnetic domain structure, function $f(\beta)$ features two pronounced maxima corresponding to magnetic fields, $\beta_1$ and $\beta_2$, in the two sorts of domains. Also, there is an interdomain boundary that appears as essentially non-zero field distribution for $\beta$ lying in the interval ($\beta_1, \beta_2$).

If we fit the actual histogram with Eq. (1) we have to deal with some criterion ($\chi^2$) defined as

$$\chi^2 = \frac{1}{N-1} \sum_{k=0}^{N-1} \frac{\left(N_k - \widetilde{N}_k\right)^2}{N_k} \tag{3}$$

that depends on many parameters. On the other hand, mainly histograms obtained in the external fields above 20kOe are interesting for study of the diamagnetic domain structure. Such fields correspond to the muon precession periods about $3.6 \cdot 10^{-9}$ s or $\sim 6\Delta t_0$ so that $\chi^2$ criterion, as a function of $\beta_1$, $\beta_2$, has many local minima due to its strong dependence on those parameters. Moreover, practically uninteresting parameter, $\varphi_0$, has to be found on that $\chi^2$ also depends strongly. These peculiarities result in rather large a number of iterations required to fit $\{\widetilde{N}_k\}$ to $\{N_k\}$. But having made a Fourier transform of experimental histogram it can be seen that useful information is given by a few harmonics whose number is much less than $N$. The other harmonics carry noise information. Dealing with the Fourier transform of the histogram makes it possible to consider only the interval of magnetic fields in that the field distribution $f(\beta)$ is of interest. Hereinafter, this interval is called "interesting" (Fig. 1). First of all, parameters $n_0$ and $b$ are found for experimental histogram $\{N_k\}$, using conventional Marquardt method [4] with fitting function

$$N_{0\,k} = n_0\, e^{-\nu\, k} + Bkg \,. \tag{4}$$

Since every $N_k$ has a Poissonian distribution, it is known [5] that

$$N_k^- \equiv \sqrt{N_k} - \sqrt{N_{0k}} \tag{5}$$

has asymptotically the Gaussian distribution with dispersion independent on $k$ and equal to $1/4$. According to the law of averages, real and imaginary parts of Fourier harmonics $\Phi_l$ of $\{N_k^-\}$ will have the normal distribution. Here, $l = 0\ldots(N/2-1)$ and since $N \gg 1$, all $\Phi_l$ will have the same dispersion $D_0$ and average of distribution $M\Phi_l = 0$. Hence, an interval of harmonics can be chosen (rather arbitrarily) to estimate the noise level $D_0$ for given histogram (Fig. 1).

Another important aspect is that one can consider not Fourier image itself but its absolute value if only the field distribution $f(\beta)$ is interesting. In this case, it is needless to find the parameter $\varphi_0$.

The matrices used in conventional Marquardt method should also be redefined. As is known [4], the matrix notation for the system of equations

$$\frac{\partial \chi^2}{\partial a_i} = 0 \qquad (6)$$

is

$$\mathbf{G} \times \Delta \mathbf{a} = \mathbf{Y} \qquad (7)$$

where $\mathbf{a}$ is the vector of parameters and $\Delta \mathbf{a}$ is its variation. Unlike [2], the matrices $\mathbf{G}$ and $\mathbf{Y}$ are given by

$$G_{ij} = \sum_{l=k_1}^{k_2} \frac{\partial |F_l|}{\partial a_i} \cdot \frac{\partial |F_l|}{\partial a_j} \qquad (8)$$

and

$$Y_i = \sum_{l=k_1}^{k_2} \left( |\Phi_l| - |F_l| \right) \cdot \frac{\partial |F_l|}{\partial a_i}, \qquad (9)$$

where partial derivatives of $|F_i|$ with respect to $a_j$ are calculated as

$$\frac{\partial |F_i|}{\partial a_j} = \frac{F_i \cdot (F_{a_j})_i^*}{|F_i|}. \qquad (10)$$

Here, $\Phi$ is the spectrum of the histogram (5), $F$ is the spectrum of the modeled histogram $\{N^-_{sqr\,k}\}$, which is defined as

$$N^-_{sqr\,k} = \sqrt{\tilde{N}_k} - \sqrt{N_{0k}}, \qquad (11)$$

$F_i$ - its $i$-th harmonic, $F_{a_j}$ is the spectrum of the modeled histogram's partial derivative with respect to the parameter $a_j$, $(\,)^*$ - complex conjugation. Interesting interval is set chosen from harmonic $k_1$ to $k_2$. Iterations themselves are performed as in conventional Marquardt method.

To describe magnetic field distribution in the sample, the following simplest model function can be used

$$f(\beta) = A_1 \cdot \delta(\beta - \beta_1) + C \cdot \theta(\beta - \beta_1) \cdot \theta(\beta_2 - \beta) + A_2 \cdot \delta(\beta_2 - \beta), \qquad (12)$$

where $\delta(\beta)$ is the delta-function of Dirac, $\theta(\beta)$ is the theta function defined as $\theta(\beta) = 0$ if $\beta < 0$ and $\theta(\beta) = 1$ if $\beta \geq 0$. But an additional broadening of Fourier maxima corresponding to the fields in domains ($\beta_1$ and $\beta_2$) was observed. That broadening could not be attributed to the mathematical peculiarities of the Fourier spectrum. To take it into account, a convolution of (12) was made with the Gaussian profile

$$f_{norm,\sigma}(\beta) = \frac{1}{\sqrt{2\pi}\sigma} \cdot \exp\left(-\frac{\beta^2}{2\sigma^2}\right), \tag{13}$$

whose $\sigma$ was also considered as an unknown parameter. The field distribution function modified in this way is given by

$$f(\beta) = f_{U,A_1}(\beta) + f_{Int,C}(\beta) + f_{U,A_2}(\beta), \tag{14}$$

where subscript $U$ denotes terms corresponding to intradomain field distribution ("uniform") and $Int$ stands for interdomain wall term. These two types of terms are expressed as

$$f_{U,A}(\beta) = \frac{A}{\sqrt{2\pi}\sigma} \cdot \exp\left(-\frac{(\beta-\beta_0)^2}{2\sigma^2}\right) \tag{15}$$

and

$$f_{Int,C} = C \cdot \int_{\beta_1}^{\beta_2} \frac{d\beta'}{\sqrt{2\pi}\sigma} \exp\left(-\frac{(\beta-\beta')^2}{2\sigma^2}\right) \tag{16}$$

Noise level is calculated as follows. A set of harmonics $\Phi_p$, where $p \in [p_1, p_2]$ ($[p_1,p_2] \cap [k_1,k_2] = \varnothing$), is chosen. First, the average value $|\Phi|_{Mean} \equiv M|\Phi_p|$ is found within the "noise interval" $[p_1, p_2]$. Then, the estimate $\sigma^2_{Fourier\,Abs}$ is found for dispersion of $|\Phi_p|$ as

$$\sigma^2_{Fourier\,Abs} = \frac{1}{p_2 - p_1} \sum_{l=p_1}^{p_2} \left(|\Phi_l| - |\Phi|_{Mean}\right)^2. \tag{17}$$

The criterion for stopping the iterations is defined as

$$\chi^2_{Fourier\,Abs} = \frac{1}{(N-1)\sigma^2_{Fourier\,Abs}} \sum_{l=0}^{N-1} \left(|F_{With\,Noise,l}| - |\Phi_l|\right)^2. \tag{18}$$

Here, $\{|F_{With\,Noise,l}|\}$ is the absolute value of the modeled histogram Fourier spectrum with the above-found noise level added according

$$|F_{With\,Noise,l}| = \sqrt{|F_l|^2 + (|\Phi|_{Mean})^2} \times f_{Elliptic}\left(\frac{|F_l| \cdot |\Phi|_{Mean}}{\sqrt{|F_l|^2 + (|\Phi|_{Mean})^2}}\right). \tag{19}$$

Here, $f_{Elliptic}(x)$ stands for polynomial approximation of the full elliptic integral $E$. To within 0.1%, $f_{Elliptic}$ is $\approx 1$.

## 3 Results

Figures 2, 3 and 4 show some visualised results of fitting. Figure 2 corresponds to uniformly magnetised sample. This fitting gave the value of Fourier maximum

broadening that was then used in processing of the domain state muon histograms. Results for these histograms are presented in Fig. 3 and 4. The first one is for external magnetic field *H=20591Oe*, the second - for *H=20596Oe*. For induction *B* in dia- and paramagnetic domains, the following values were obtained: 20574*Oe* and 20605Oe for Fig. 3, 20579*Oe* and 20607*Oe* for Fig. 4. For the histograms analysed, a change of Fourier harmonic number by unity was equivalent to the change of magnetic field *B* by $\Delta B = 7.2063\,Oe$. It can be seen that the values of magnetic fields in domains remain practically constant over the domain state interval of external field. The volume fraction of interdomain boundary decreased from 55% to 40%, also there was a decrease of diamagnetic phase volume fraction from 17% to 13%, whereas the volume fraction of paramagnetic phase increased from 27% to 47%.

It was interesting to estimate the precision of the calculated interdomain boundary volume fraction. For that, the following was done. First, the Eq.1 with the field distribution (14) was used to create a simulated histogram. The distribution parameters were taken from the fitting that was made previously for the experimental histogram. Then the Poissonian noise was introduced to the simulated histogram using a pseudo-random number generator. The last step was repeated 20 times to get a representative set of simulated histograms with *a priori* the same parameters. Then the fitting procedure was applied to those histograms. It resulted in 20 sets of field distribution parameters ($A_1$, $\beta_1$, $A_2$, $\beta_2$ and $C$). Estimated accuracy ($1\sigma$) for $A_1$, $\beta_1$, $A_2$, $\beta_2$ and $C$ amounted to 14%, 10%, 0.004%, 0.003% and 10% correspondingly.

## 4 Discussion

The question of the domain structure period was first considered by Privorotskii [6]. He showed that

$$X \sim \sqrt{ZD} \tag{20}$$

Here, *X* is the period of structure, *Z* stands for the sample thickness, *D* is the Larmor diameter taken as an estimate of the interdomain boundary thickness. An improved estimate made by Condon is given in [8]

$$X = \alpha\sqrt{ZD} \tag{21}$$

Dimensionless parameter $\alpha$ is weakly dependent on magnetic interaction strength $a = 4\pi|dM/dB|$. Condon shows that $\alpha$ increases with *a* as follows: $\alpha = 0$ for $a = 1$, $\alpha = 2$ for $a = 4$ and $\alpha = 3.5$ for $a = 10$. Assuming $D \sim 10^{-4}$ cm (for Be in $H \sim 20\,\text{kG}$), $Z \sim 10^{-1}$ cm, $a \sim 3$ (and, consequently, $\alpha \sim 2$) one can obtain $X \sim 10^{-2}$ cm. As a result, the boundary volume fraction, given by (21) is about 1% that is not in line with what the processing of muon histograms gives. This discrepancy might be due to not taking into account the magnetostriction of the sample. In domain state, the diamagnetic phase contracts while the paramagnetic one expands. The interdomain boundaries experience shear deformations since they act as mediators between those two sorts of domains with opposite character of magnetostriction. This deformation results in a positive contribution to the free energy of the sample. It can be shown that the value of this contribution increases with $Z$ and with decreasing $D$. Hence, the assumption that $D$ is independent on $Z$ is nottrue at least for rather thick samples.

## 5 Conclusion
The use of modified Marquardt minimisation procedure, makes it possible to consider muon histograms whose field distribution contains two closely spaced maxima with some non-zero value between them. Quantitative parameters of some muon histograms were determined. The interdomain boundary volume fraction turned out to be up to ~50%.

Author is grateful to Prof. E.L.Kosarev, Dr. V.S.Egorov and Prof. E.P.Krasnoperov for useful discussion and corrections.

*Philip Lykov RRC "Kurchatov Institute" Study of interdomain boundary*

**Figures**

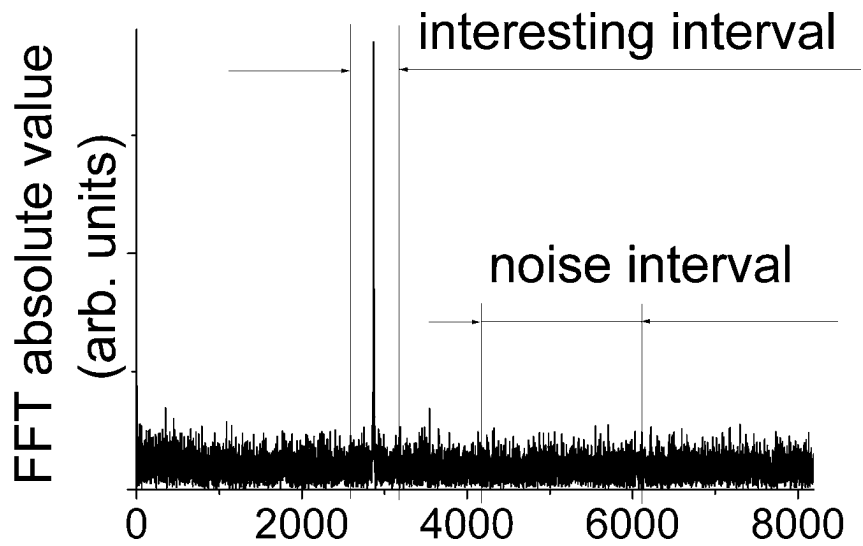

Figure 1. "Interesting" interval and "noise" interval as they are specified within the histogram Fourier harmonics set.

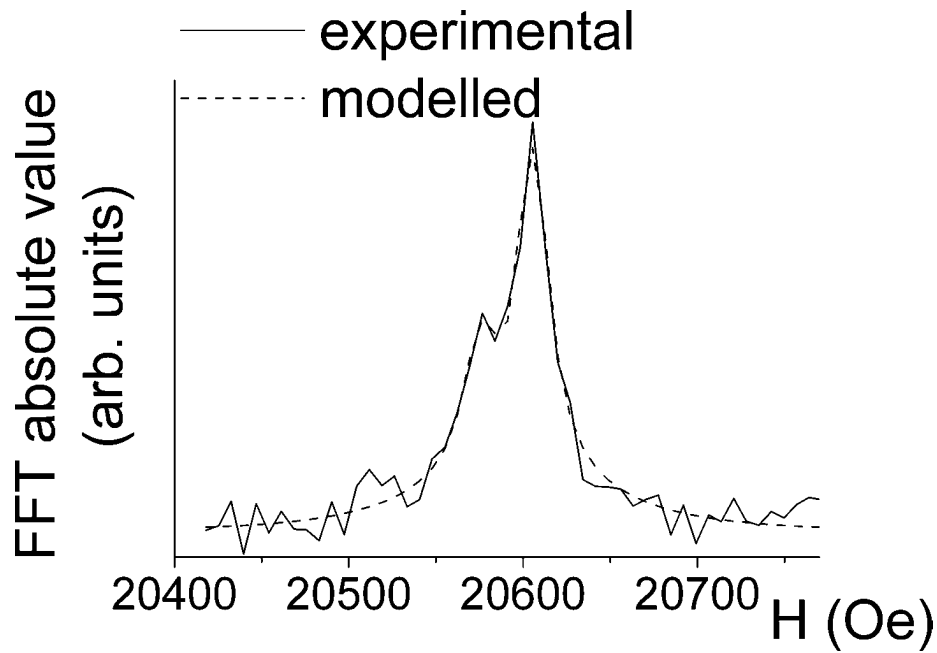

Figure 2. "Interesting" interval of uniformly magnetised sample histogram FFT image. Histogram is analysed in order to obtain field broadening, $\delta B$, that determines $\sigma$ value. Black line - FFT of experimental data, grey line - the result of the fitting.

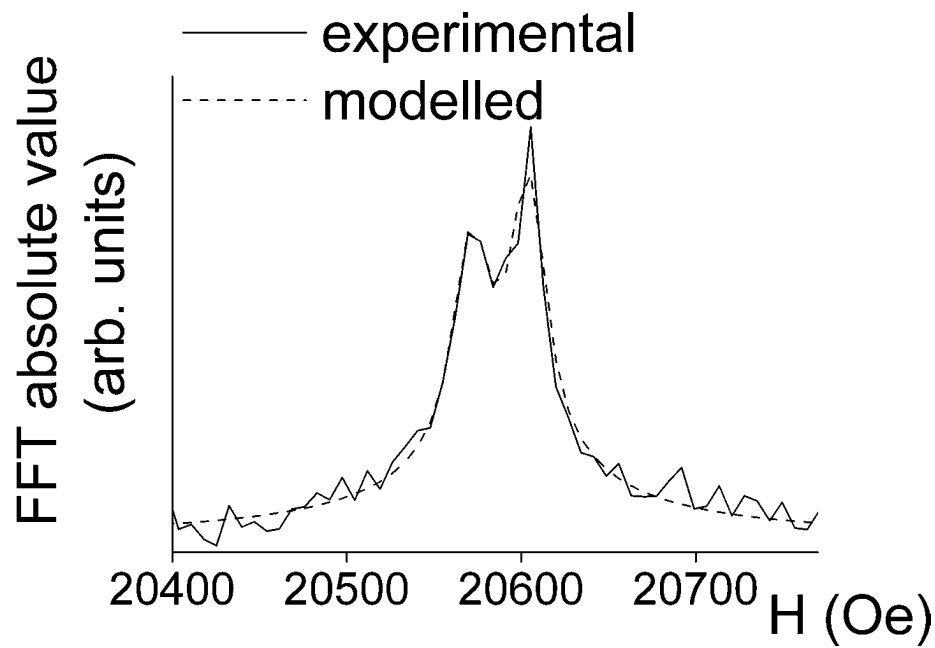

Figure3. "Interesting" interval for domain state muon histogram obtained in external magnetic field 20591*Oe* volume fractions of diamagnetic/ interdomain/ paramagnetic phases are 17/56/27% correspondingly.

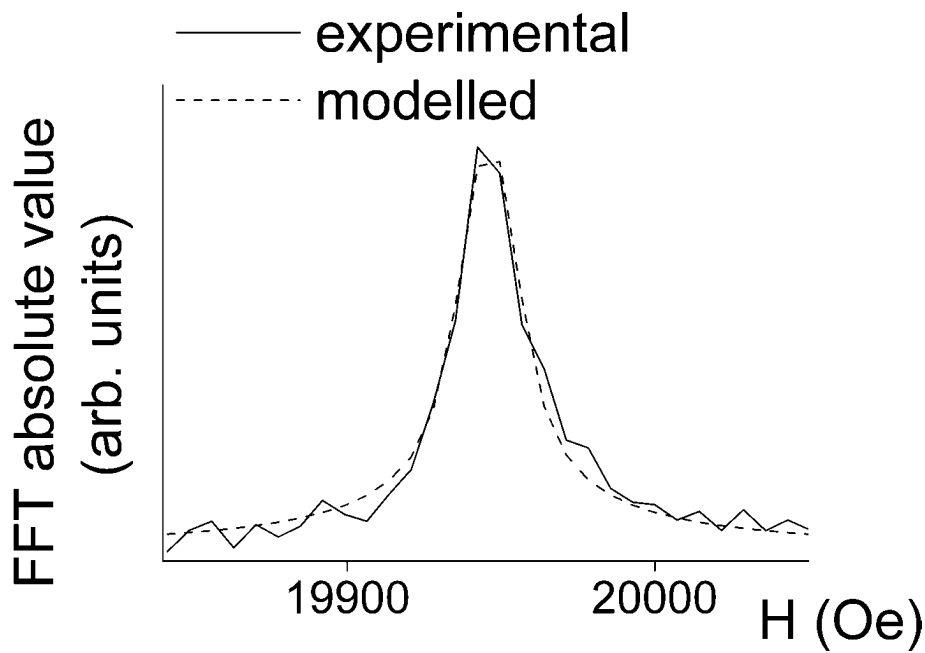

Figure 4. The same as Figure 3 but for external field 20596*Oe*; volume fractions are 13/40/47%.